\documentclass[twocolumn,showpacs,preprintnumbers,amsmath,amssymb]{revtex4-1}

\usepackage{bbm}
\usepackage[caption=false]{subfig}
\usepackage{overpic}
\usepackage{amsmath}
\usepackage{epsfig,psfrag}
\usepackage{pstricks}
\usepackage{dcolumn}
\usepackage{bm}
\usepackage{graphicx}
\usepackage{color}
\usepackage{version}
\newcommand{\mc}[1]{\textcolor{red}{{\it\textbf{#1}}}}

\usepackage{times}

\begin{document}
\title{The Interacting Mesoscopic Capacitor Out of Equilibrium}
\author{Daniel Litinski$^1$}
\author{Piet W. Brouwer$^1$}
\author{Michele Filippone$^{1,2}$}
\affiliation{$^1$Dahlem Center for Complex Quantum Systems and Institut f\"ur Theoretische Physik, Freie Universit\"at Berlin, Arnimallee 14, 14195 Berlin, Germany}
\affiliation{$^2$DQMP, University of Geneva, 24 Quai Ernest Ansermet, 1211 Geneva, Switzerland}

\begin{abstract}
We consider the full non-equilibrium response of a mesoscopic capacitor in the large transparency limit, exactly solving a model with electron-electron interactions appropriate for a cavity in the quantum Hall regime. For a cavity coupled to the electron reservoir via an ideal point contact, we show that the response to any time-dependent gate voltage $V_{\rm g}(t)$ is strictly linear in $V_{\rm g}$. We analyze the charge and current response to a sudden gate voltage shift, and find that this response is not captured by a simple circuit analogy. In particular, in the limit of strong interactions a sudden change in the gate voltage leads to the emission of a sequence of multiple charge pulses, the width and separation of which are controlled by the charge-relaxation time $\tau_{\rm c} = h C_{\rm g}/e^2$ and the time of flight $\tau_{\rm f}$. We also consider the effect of a finite reflection amplitude in the point contact, which leads to non-linear-in-gate-voltage corrections to the charge and current response. 
\end{abstract}

\maketitle

\section{Introduction}\label{sec:intro}

\begin{figure}
	\centering
	\includegraphics[width=\linewidth]{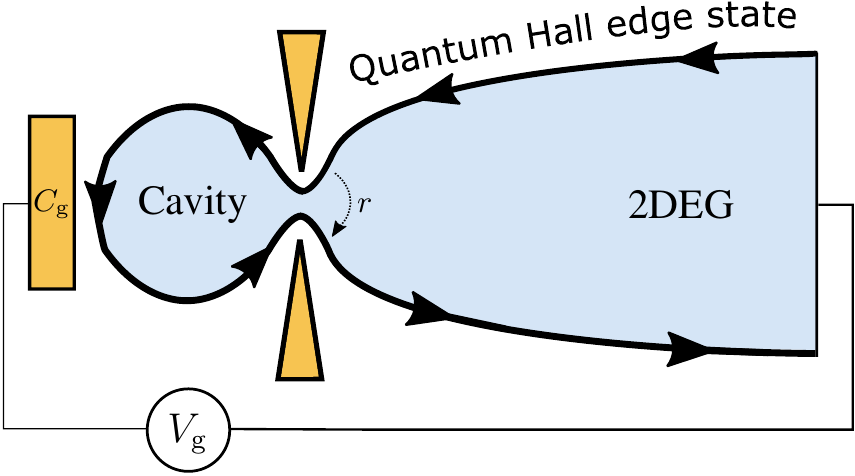}
	\caption{The mesoscopic capacitor realized with a quantum cavity coupled to the quantum Hall edge state of a bulk two-dimensional electron gas through a quantum point contact. A capacitor plate with geometrical capacitance $C_{\rm g}$ and voltage $V_{\rm g}$ controls the charge of the cavity. The backscattering amplitude $r$ of electrons at the cavity entrance is controlled by the opening of the quantum point contact. }\label{fig:setup}
\end{figure}

The mesoscopic capacitor has played a central role in the quest to achieve full control of scalable coherent quantum systems \cite{loss1998,petta2005coherent,koppens2006driven}. A mesoscopic capacitor is an electron cavity (quantum dot) coupled to a lead via a quantum point contact and capacitively coupled to a metallic gate \cite{buttiker1993,buttiker1993b,buttiker1996}. The interest in this device stems from the absence of dc transport, which makes the direct investigation and control of the coherent dynamics of charge carriers possible. The first experimental realization of this system by Gabelli {\it et al.} consisted of a two-dimensional ``cavity'' in the quantum Hall regime \cite{gabelli2006,gabelli12}, the ``lead'' being the edge of a bulk two-dimensional electron gas, see Fig.~\ref{fig:setup}. Operated out of equilibrium and in the weak tunneling limit, this system allows the triggered emission of single electrons \cite{feve2007,mahe2010,parmentier2012} and has paved the way to the realization of quantum optics experiments with electrons \cite{bocquillon2012,bocquillon2013,bocquillon2014,grenier11},   as well as probing electron fractionalization \cite{bocquillon13separation,freulon15} and relaxation \cite{marguerite16}. On-demand single-electron sources were also recently realized relying on real-time switching of tunnel-barriers \cite{leicht11,battista11,fletcher13,waldie15,kataoka16,johnson16}, ``electron sound-wave surfing'' \cite{hermelin11,bertrand15,bertrand16}, the generation of levitons \cite{levitov96,ivanov97,keeling2006,dubois13,rech16} and superconducting turnstiles \cite{vanzanten16,basko17}.

The key fundamental questions related to the dynamics of a mesoscopic capacitor  are about the relaxation of its charge $Q$ following a change in the gate voltage $V_{\rm g}$ and the electronic state subsequently emitted from the cavity. The linear response is characterized by the ``admittance'' ${\cal A}(\omega)$,
\begin{equation}
  Q(\omega) = {\cal A}(\omega) V_{\rm g}(\omega) +
  {\cal O}(V_{\rm g}^2).
  \label{eq:admittance}
\end{equation}
In their seminal work, B\"uttiker and coworkers showed that the low-frequency admittance of a mesoscopic capacitor has the form of the admittance of a classical $RC$ circuit \cite{buttiker1993,buttiker1993b,buttiker1996},
\begin{equation}
  {\cal A}(\omega)= C (1+ i\omega R_{\rm q} C) + \mathcal{O}(\omega^2),
\label{eqn:rccircuit2}
\end{equation}
with a charge relaxation resistance $R_{\rm q} = h/2 e^2$ universally equal to half of the resistance quantum \cite{gabelli2006}, independent of the transparency of the quantum point contact connecting the cavity to the lead. The expansion \eqref{eqn:rccircuit2} also applies in the presence of interactions in the cavity \cite{nigg2006,ringel2008,mora2010,dutt2013strongly}
and the universality of the charge relaxation resistance $R_{\rm q}$ was shown to have its roots in a Korringa-Shiba relation \cite{filippone2012}. Deviations from universality arise in non-Fermi liquid regimes \cite{hamamoto2010,hamamoto2011quantum,mora13,burmistrov15}, or for the low-temperture limit of an Anderson impurity, upon breaking the Kondo singlet by an applied magnetic field \cite{lee2011,filippone2011,filippone2013}. An effective $RC$ circuit also plays a central role in the photon-charge interaction in novel quantum hybrid circuits \cite{delbecq11,delbecq13,schiro14,liu14,bruhat16,mi16} and in energy transfer \cite{ludovico14,romero16}. 

The circuit analogy \eqref{eqn:rccircuit2} does not apply for a non-linear response to a gate voltage change or to fast (high-frequency) drives. An important example is a large step-like change in the gate voltage $V_{\rm g}(t)=V_{\rm g}\theta(t)$, $\theta(t)$ being the Heaviside step function, which is relevant to achieve triggered emission of quantized charge \cite{parmentier2012}. Such non-linear high-frequency response has been considered extensively for non-interacting cavities \cite{feve2007,parmentier2012,keeling2008,Olkhovskaya08,moskalets2008,sasaoka2010,moskalets2013}, where the current response to a gate voltage step at time $t=0$ was found to be of the form of simple exponential relaxation \cite{feve2007,keeling2008,moskalets2008,moskalets2013}
\begin{equation}\label{eq:iscatt}
  I(t)\propto e^{-t /\tau_{\rm R}} \theta(t).
\end{equation}
For a cavity in the quantum Hall regime the relaxation time $\tau_{\rm R} = \tau_{\rm f}/(1-|r|^2)$, where $\tau_{\rm f}$ is the time of flight around the edge state of the cavity, see Fig.~\ref{fig:setup}, and $r$ the reflection amplitude of the point contact.

There have been relatively few studies of the out-of-equilibrium behavior of the mesoscopic capacitor in the presence of interactions. The dominant electron-electron interactions in the cavity have the form of a charging energy \cite{grabert93,aleiner2002}
\begin{equation}\label{eq:charging}
  H_{\rm int} = \frac{e^2}{2 C_{\rm g}} [N-{\cal N}_{\rm g}(t)]^2\,,
\end{equation}
in which $N$ is the number of electrons in the cavity, $C_{\rm g}$ the geometric capacitance, and ${\cal N}_{\rm g} = C_{\rm g} V_{\rm g}(t)/e$ is the dimensionless gate voltage \footnote{Relying on a model in which long-range interactions are considered only within the cavity is justified by the confinement of the chiral edge within a small region such as the cavity. The outer edges are considerably spatially separated, in particular in experiments such as those reported in Refs.~\cite{feve2007,mahe2010,parmentier2012} they should not interact with each other. Nevertheless, in Appendix~\ref{app:finiteswitching}, we discuss how corrections to this assumption are readily included in our treatment of the problem, without modifying the essence of our results.}. The charging energy leads to an additional time scale $\tau_{\rm c} = 2 \pi \hbar C_{\rm g}/e^2$ for charge relaxation. The limit $1-|r|^2 \ll 1$ of a cavity weakly coupled to the lead, such that it can effectively be described by a single level, was addressed in Refs.\ \cite{splett2010,contreras2012,kashuba12,alomar15,alomar16,vanherck16}. An important partial result for the opposite limit of an almost transparent point contact was obtained by Mora and Le Hur \cite{mora2010}, who studied the linear response without restriction on the frequency $\omega$, for a cavity in the quantum Hall regime. Their result for the admittance ${\cal A}(\omega)$ for a fully transparent point contact ($r=0$),
\begin{equation}\label{eqn:rccircuit1}
  {\cal A}(\omega) = C_{\rm g} \left(1- \frac{i\omega\tau_{\rm c}}{1-e^{ i  \omega\tau_{\rm f}}}\right)^{-1}\,,
\end{equation}
features both time scales $\tau_{\rm f}$ and $\tau_{\rm c}$, leading to a charge relaxation behavior considerably more complicated than that of Eq.\ (\ref{eq:iscatt}), although the two time scales $\tau_{\rm f}$ and $\tau_{\rm c}$ still combine into the universal charge relaxation resistance $R_{\rm q} = h/2e^2$ and capacitance \cite{buttiker1993,buttiker1993b,buttiker1996}
\begin{equation}
  \frac 1C = \left[ \frac{1}{C_{\rm g}} + \frac{h}{e^2 \tau_{\rm f}} \right]
  \label{eq:C}
\end{equation}
upon expanding Eq.\ (\ref{eqn:rccircuit1}) to linear order in $\omega$.  The admittance \eqref{eqn:rccircuit1} also appears in the description of the coherent transmission of
electrons through interacting Mach-Zehnder interferometers \cite{dinh12,dinh13}.

In this work, we report a study of the full out-of-equilibrium behavior of the mesoscopic capacitor with a close-to-transparent point contact, thus extending the calculation of Ref.\ \onlinecite{mora2010} to non-linear response in the gate voltage $V_{\rm g}$. As in Ref.\ \onlinecite{mora2010} we consider a cavity in the quantum Hall regime, so that the time scale $\tau_{\rm f}$ can be identified with the propagation time along the cavity's edge. A main result, spectacular in its simplicity, is that for a fully transparent contact ($r=0$) the linear-response admittance (\ref{eqn:rccircuit1}) also describes the non-linear response, {\it i.e.}, the correction terms in Eq.\ (\ref{eq:admittance}) vanish for an ideal point contact connecting cavity and lead \cite{cuniberti98}. Further, we analyze the charge evolution after a step change in the gate voltage and show that initially, for times up to $\tau_{\rm f}$, $Q(t)$ relaxes exponentially with time $\tau_{\rm c}$, whereas at time $t = \tau_{\rm f}$ the capacitor abruptly enters a regime of exponentially damped oscillations, the period and the exponential decay of which are controlled by a complex function of $\tau_{\rm f}$ and $\tau_{\rm c}$, which does not correspond to any time scale extracted from low-frequency circuit analogies. This behavior is not captured by Eq.~\eqref{eq:iscatt}, derived in the non-interacting limit. We show that these oscillations correspond to the emission of initially sharp charge density pulses, which are damped and become increasingly wider after every charge oscillation.  Finally, we also consider the effect of a small reflection amplitude $r$ in the point contact, where we do find that the charge $Q_r$ acquires nonlinear terms in the gate voltage $V_{\rm g}$,
\begin{equation}\label{eqn:backscattering5}
  Q_r(t)= Q(t) - \frac{e \tilde{r}}{\pi C} \int dt'\,  \mathcal A(t-t') \sin[2 \pi Q(t')/e],
\end{equation}
in which $\mathcal A(t)$ and $Q(t)$ are the Fourier transform of the admittance and charge for the case of a point contact with perfect transparency, $r=0$, see Eqs.\ \eqref{eq:admittance} and (\ref{eqn:rccircuit1}). The parameter $\tilde r$ involves both the (weak) backscattering amplitude $r$ and temperature $T$, and can be found in Eq. \eqref{eq:erre} below.

Our calculation employs the bosonization formalism \cite{haldane81,haldane81prl,giamarchi2004quantum} to map interacting fermions to non-interacting bosons \cite{matveev1995,aleiner1998b,brouwer2005}. For a transparent point contact, the bosononization formalism allows to derive {\it exact} results for the out-of-equilibrium behavior of the interacting mesoscopic capacitor. The only approximation is that the point contact's transparency remains perfect for all energies of interest. This is not a serious limitation, since the latter energy range is independent of the cavity size, whereas the typical energy scales $\hbar/\tau_{\rm c}$ and $\hbar/\tau_{\rm f}$ for the capacitor's response go to zero in the limit of a large cavity size. Given the microscopic nature of our approach, we describe the propagation of charge pulses within the cavity edge, a study which is complementary to that of electron waiting times of the cavity \cite{mathias2011,mathias2012,hofer16}. Moreover our approach avoids the mean-field approximation underlying scattering theory approaches \cite{buttiker1993,buttiker1993b,buttiker1996,feve2007,keeling2008,moskalets2008,parmentier2012,moskalets2013,moskalets11}, and shows how interactions trigger remarkable and novel coherence effects.

The remainder of this article is structured as follows: In Sec.~\ref{sec:model}, we introduce the specific model of an interacting mesoscopic capacitor in the quantum Hall regime and describe its formulation in the bosonization formalism. In Sec.~\ref{sec:opencavity}, we solve the model for the case of a point contact with perfect transparency and show that Eq.\ (\ref{eqn:rccircuit1}) also describes the non-linear response to a gate voltage $V_{\rm g}$. In Sec.~\ref{sec:step}, we examine in detail the charge response of the cavity to a step change of the gate voltage and compare it to the ``$RC$'' \cite{buttiker1993,buttiker1993b,buttiker1996,mora2010} and ``$RLC$'' \cite{wang2007,yin2014} circuit analogies. In Sec.~\ref{sec:pulses}, we consider the propagation of charge pulses along the cavity edge and show how these lead to the serial emission of multiple charge density pulses from the cavity. In Sec.~\ref{sec:backscattering}, we consider the effect of a nonideal quantum point contact to first order in the backscattering amplitude. We derive Eq.\ (\ref{eqn:backscattering5}) and investigate how backscattering affects the emission of charge pulses. We conclude with a brief outlook in Sec.\ \ref{sec:conclusion}.


\section{Model}\label{sec:model}

\begin{figure}
	\centering
	\includegraphics[width=\linewidth]{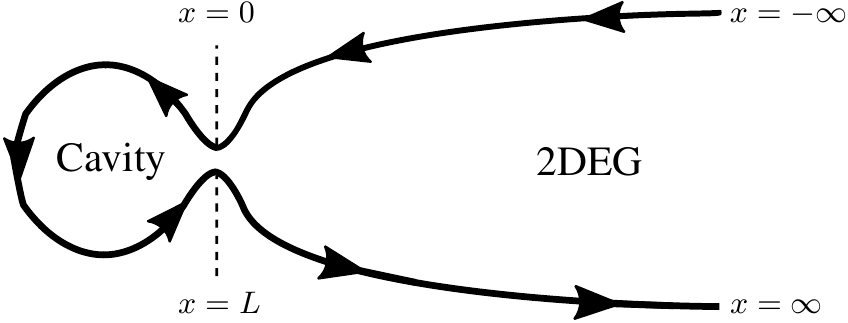}
	\caption{Schematic picture of the mesoscopic capacitor in the quantum Hall regime. The coordinate $x$ labels the electron position along the one-dimensional chiral edge state shown in the figure. The point of entrance into the cavity is labeled ``$x=0$''. The exit from the cavity is at ``$x=L$'', with $L = v \tau_{\rm f}$.} \label{fig:model}
\end{figure}

We consider a mesoscopic capacitor consisting of a cavity and the adjacent bulk two-dimensional electron gas in a large perpendicular magnetic field, so that cavity and bulk are in the quantum Hall regime. The system is shown schematically in Fig.\ \ref{fig:model}. 

The relevant electronic degree of freedom is the one-dimensional chiral edge state of the cavity and the bulk two-dimensional electron gas, which can be described by a single propagating chiral mode that propagates along the edge of the bulk two-dimensional electron gas, passes through the cavity, and continues along the edge of the bulk electron gas \cite{fabrizio1995,mora2010}. For simplicity we assume that the propagation velocity $v$ along the chiral edge is constant. We use the coordinate $x$ to label the position along the edge, and choose $x=0$ to be the point of entrance into the cavity, see Fig.\ \ref{fig:model}. The second passage through the quantum point contact, upon exiting the cavity, then is at $x = L$, with
\begin{equation}
  L = v \tau_{\rm f}.
\end{equation}
Together with the interaction (\ref{eq:charging}) this gives the Hamiltonian
\begin{equation}
  H = -i \hbar v \int dx\, \psi^\dagger(x) \partial_x \psi(x) 
  + \frac{e^2}{2 C_{\rm g}} [N - {\cal N}_{\rm g}(t)]^2,
  \label{eq:H}
\end{equation}
where $\psi^{\dagger}(x)$ and $\psi(x)$ are the creation and annihilation operators for an electron at the chiral edge and the particle number 
\begin{equation}
  N = \int_0^L dx\, :\! \psi^{\dagger}(x) \psi(x)\!:\,.
\end{equation}
Backscattering at the cavity entrance, with reflection amplitude $r$, is described by an additional term
\begin{equation}
  H_r = -\hbar r v [\psi^{\dagger}(L) \psi(0) + \psi^{\dagger}(0) \psi(L)].
\end{equation}

The interacting Hamiltonian (\ref{eq:H}) can be brought to quadratic form using the bosonization identities \cite{vondelft1998,kane2005,giamarchi2004quantum} 
\begin{equation}\label{eq:bos}
 :\!\psi^{\dagger}(x) \psi(x)\!:\, = \frac{\partial_x \phi(x)}{2\pi},\ \
	\psi^\dagger(x) = \frac{1}{\sqrt{2\pi a}}e^{i\phi(x)}\,,
\end{equation}
in which $\phi(x)$ is a real bosonic field obeying the Kac-Moody relation $[\partial_x\phi(x),\phi(x')]=2\pi i\delta (x-x')$ and $a$ is a short-distance cutoff. Applying Eq.~\eqref{eq:bos}, the number of particles in the cavity $N$ becomes linear in bosonic fields,
\begin{equation}
  N = \frac{1}{2 \pi}[\phi(L)-\phi(0)],
  \label{eq:N}
\end{equation}
and, hence, the charging term in \eqref{eq:H} becomes quadratic in $\phi$ \cite{matveev1995}. This gives the bosonized Hamiltonian
\begin{equation}\label{eq:free}
  H = \frac{\hbar v}{4\pi} \int\limits_{-\infty}^{\infty} \mathrm{d}x ~ [\partial_x \phi(x)]^2 + \frac{e^2}{2 C_{\rm g}} \left( N - {\cal N}_{\rm g} \right)^2,
\end{equation}
where $N$ is expressed in terms of the bosonic field $\phi(x)$ as in Eq.\ (\ref{eq:N}). The backscattering term $H_r$, which was quadratic in the fermionic fields $\psi(x)$, is no longer quadratic in the bosonic formulation,
\begin{equation}\label{eq:bs}
  H_r=-\frac {\hbar v r}{\pi a}\cos(2 \pi N)\,.
\end{equation}
In the fully transparent limit, $r=0$ and Eq.~\eqref{eq:free} is exactly solvable for any drive $V_{\rm g}(t) = e {\cal N}_{\rm g}(t)/C_{\rm g}$, as we will discuss in the next Section. The effects of finite backscattering will be addressed to first order in perturbation theory in $H_r$ in Sec.~\ref{sec:backscattering}.


\section{Open cavity}\label{sec:opencavity}

For a perfectly transparent contact of the mesoscopic capacitor, the interacting system is described by Eq.\ \eqref{eq:free}. Since this is a quadratic Hamiltonian, we can find an exact solution for the field $\phi(x,t)$ for an arbitrary time-dependent voltage $V_{\rm g}(t) = e {\cal N}_{\rm g}(t)/C_{\rm g}$. The Heisenberg equation of motion for the fields $\phi(x,t)$ reads 
\begin{equation}\label{eq:eqm}
  \frac{\partial \phi(x,t)}{\partial t} =-v_F \frac{\partial \phi(x,t)}{\partial x} -
  \frac{2 \pi s(x)}{\tau_{\rm c}}
  \left[N(t)-{\cal N}_{\rm g}(t)\right],
\end{equation}
with $\tau_{\rm c} = 2 \pi \hbar C_{\rm g}/e^2$ and
\begin{equation}\label{eq:s}
  s(x) = \left\{ \begin{array}{ll} 1 & \mbox{if $0 < x < L$},\\
  0 & \mbox{else}. \end{array} \right.
\end{equation} 
The field $\phi(0)$ at the entrance to the mesoscopic capacitor is unaffected by the interaction term and the time-dependent gate voltage ${\cal N}_{\rm g}(t)$, so that all its correlation functions are those of a {\it free} bosonic field {\it in equilibrium}. Using Eq.\ (\ref{eq:eqm}) the field $\phi(x)$ for $x > 0$ and the charge field $N$ can be expressed in terms of the field $\phi(0)$ at the cavity entrance. Hereto, we first use a direct solution of Eq.\ (\ref{eq:eqm}) for $\phi(x)$ at arbitrary $x > 0$ in terms of $\phi(0)$ and $N$,
\begin{align}
  \phi(x,t) =&\, \phi(0,t-x/v) 
  \label{eq:sol}
  \\ &\, \mbox{} - \frac{2 \pi}{\tau_{\rm c}} \int^t dt'
  \left[ N(t') - {\cal N}_{\rm g}(t') \right]
  s[x-v (t-t')], \nonumber 
\end{align}
and then use Eq.\ (\ref{eq:N}) in combination with Eq.\ (\ref{eq:sol}) for \linebreak $x=L$ to express $N$ in terms of $\phi(0)$ (see App.\ \ref{app:backscattering} for details),
\begin{align}
  N(t) =&\, \int^t dt' {\cal A}(t-t')
  \left[ \frac{{\cal N}_{\rm g}(t')}{C_{\rm g}} 
  + \frac{\hbar}{e^2} \frac{\partial \phi(0,t')}{\partial t'}
  \right],
  \label{eq:phi0eq}
\end{align}
where ${\cal A}(t-t')$ is the Fourier transform of the admittance (\ref{eqn:rccircuit1}) found previously by Mora and Le Hur \cite{mora2010}. Upon using $\langle Q(t) \rangle = e \langle N(t) \rangle$ and $\langle \phi(0,t) \rangle = 0$, Eq.\ (\ref{eq:phi0eq}) immediately reproduces the Fourier transform of Eq.\ (\ref{eqn:rccircuit1}),
\begin{equation}
  \langle Q(t) \rangle =\int^t dt' \mathcal A(t-t')V_{\rm g}(t').
  \label{eq:drive}
\end{equation}  

Although Eq.\ (\ref{eq:drive}) formally coincides with the result previously obtained in Ref.\ \onlinecite{mora2010} for the linear response to a gate voltage change, the present derivation makes no assumption regarding the magnitude of $V_{\rm g}(t)$ and, hence, shows that the charge response of the open cavity is {\it always} linear in the gate voltage $V_{\rm g}(t)$, no matter how strong or fast its variations are.

The reason that the linear behavior extends to arbitrary strengths of
the driving voltage is that in the absence of backscattering in the
point contact, the charge-density (``plasmonic'') excitations of the
chiral edge are non interacting objects which couple linearly to the
gate voltage $V_{\rm g}(t)$. In our formalism, these two aspects are
responsible for the linear dependence of the field operator $\phi(x,t)$
on the gate voltage $V_{\rm g}(t)$, as given explicitly by the
Heisenberg equation of motion \eqref{eq:eqm} and its exact solution \eqref{eq:sol}.
Alternatively, the reason for this linear behavior can be attributed to the  complete delocalization of  the eigenmodes of the open-cavity system. Thus, any local perturbation couples to an infinity of delocalized modes, and therefore even strong perturbation turn into small, close-to-equilibrium
perturbations on each eigenstate. As we will discuss in
Sec. VI, a finite backscattering amplitude leads to state localization within the cavity and then non-
linear corrections to the charge response, see Eq.~\eqref{eqn:backscattering5}. This argument applies in the absence of interactions as well. 

Equation \eqref{eqn:rccircuit1} for the admittance ${\cal A}(\omega)$ clearly shows the existence of two time scales affecting the dynamics of the mesoscopic capacitor: the time scale $\tau_{\rm c}$ for charge relaxation of the cavity, and the time $\tau_{\rm f}$ that charge density excitations require to travel along the cavity edge. The low-frequency expansion of Eq.~\eqref{eqn:rccircuit1} reproduces the $RC$ form of Eq.~\eqref{eqn:rccircuit2} and combines $\tau_{\rm c}$ and $\tau_{\rm f}$ in a single $RC$ time
\begin{equation}\label{eq:trrcshort}
\tau_{RC} = R_{\rm q} C = \frac{1/2}{1/\tau_{\rm c}+1/\tau_{\rm f}}\,.
\end{equation} 
If the large-cavity limit $\tau_{\rm f} \rightarrow\infty$ is performed before the $\omega\rightarrow 0$ expansion in Eq. \eqref{eqn:rccircuit1}, the $RC$ time crosses over to $\tau_{\rm c}$ \cite{mora2010}, which differs from the $RC$ time one obtains from Eq.\ (\ref{eq:trrcshort}) by taking the limit $\tau_{\rm f} \to \infty$ after taking the zero-frequency limit. The separate roles of $\tau_{\rm c}$ and $\tau_{\rm f}$ emerge only in the non-adiabatic setting, {\it i.e.}, considering the response at finite frequency or the response to sudden changes of the gate voltage. 

To obtain the real-time response function ${\cal A}(t-t')$, the Fourier transform of the admittance (\ref{eqn:rccircuit1}) has to be calculated, which can be accomplished via standard complex contour integration. All poles of $\mathcal A(\omega)$ lie in the lower complex plane and are given by   
\begin{equation}\label{eq:pole}
\mathcal \omega_n =-\frac i{\tau_{\rm c}} f_n(\tau_{\rm f}/\tau_{\rm c}),\ \ n\neq 0\,,
\end{equation}
in which
\begin{equation}\label{eq:fn}
f_n(\tau_{\rm f}/\tau_{\rm c}) = 1-\frac{\tau_{\rm c}}{\tau_{\rm f}}W_n(\tau_{\rm f} e^{\tau_{\rm f}/\tau_{\rm c}}/\tau_{\rm c})
\end{equation}
and $W_n(z)$ is the $n$-th branch of the Lambert $W$ function (also called product logarithm). This gives the following expression for the admittance ${\cal A}(t)$,
\begin{equation}\label{eq:at}
  {\cal A}(t)=\frac{e^2}h\theta(t)\sum\limits_{n\neq 0} \frac{\tau_{\rm c} f_ne^{-f_nt/\tau_{\rm c}}}{\tau_{\rm c} + \tau_{\rm f} - \tau_{\rm f} f_n},
\end{equation}
where we suppressed the argument of the function $f_n$ defined in Eq.\ (\ref{eq:fn}) above.
The charge and current response described by this function will be investigated in detail in the next two Sections.
In Appendix \ref{app:lambert}, we summarize relevant properties of these functions, in particular that $f_n = f_{-n}^*$ and that, for any fixed $\tau_{\rm f}/\tau_{\rm c}$, the real part  of $f_n$ is positive and increases with increasing $n$.


\section{Response to a step voltage}\label{sec:step}

We now investigate in detail the charge response to a sudden modification of the gate voltage, $V_{\rm g}(t)= V_{\rm g}(0) + \Delta V_{\rm g}\theta(t)$. Substituting Eq.\ (\ref{eq:at}) into Eq.\ (\ref{eq:drive}) we then find that the change $\Delta Q(t) = Q(t) - Q(0)$ of the charge on the cavity is, for $t > 0$,
\begin{equation}\label{eqn:opencavity3}
  \Delta Q(t) = C \Delta V_{\rm g}
  -\sum\limits_{n\neq 0} \frac{C_{\rm g} \tau_{\rm c}
  \Delta V_{\rm g} e^{-f_n t/\tau_{\rm c}}}{
  \tau_{\rm c} + \tau_{\rm f}- \tau_{\rm f} f_n},
\end{equation}
where the first term is the equilibrium charge response to a gate voltage change $\Delta V_{\rm g}$, which involves the total capacitance $C^{-1} = C_{\rm g}^{-1} + h/e^2 \tau_{\rm f}$, see Eq.\ (\ref{eq:C}).
The contribution $h/e^2 \tau_{\rm f}$ to the inverse capacitance is usually understood as a manifestation of the additional energy required by Pauli exclusion principle to add electrons in the cavity \footnote{The total capacitance of the systems is just given by the charge susceptibility $\partial Q/\partial V_{\rm g}$, which is the density of `charge' states in the cavity. This distinction is important when charge and other electron degrees of freedom decouple because of interactions, a paradigmic  scenario being the presence of Kondo correlations, see Ref. \cite{filippone2012} for a detailed discussion. An alternative interpretation of this ``quantum capacitance'' based on dynamic arguments naturally appears in our approach and will be discussed in Sec. \ref{sec:pulses}.}. Some subtleties concerning the derivation of Eq.\ (\ref{eqn:opencavity3}) are discussed in Appendix \ref{app:numeval}. 

\begin{figure}[t]
	\centering
	\includegraphics[width=\linewidth]{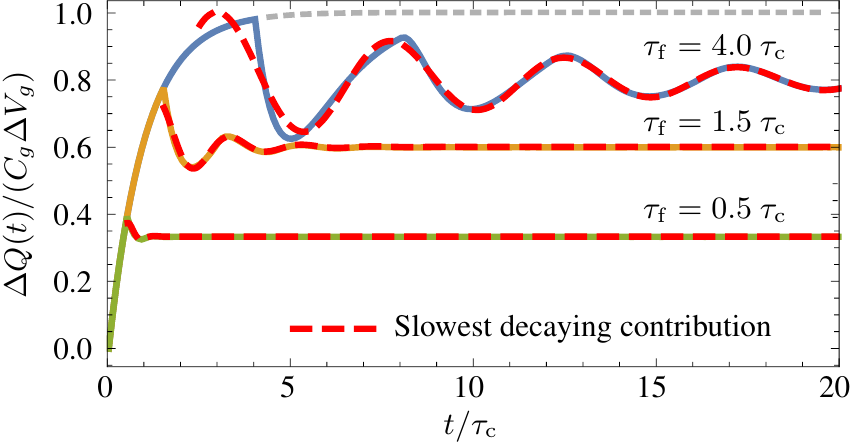}
	\caption{Change $\Delta Q(t)$ of the cavity charge after a step change in the gate voltage $V_{\rm g}(t)=\theta(t) \Delta V_{\rm g}$ for a mesoscopic capacitor with a fully transparent point contact ($r=0$). Different curves correspond to  different interaction strengths, characterized by the ratio $\tau_{\rm f}/\tau_{\rm c}$. For $t < \tau_{\rm f}$ the charge evolution equals that of an $RC$ circuit with $RC$ time $h C_{\rm g}/e^2$ (grey dotted line). At $t=\tau_{\rm f}$, damped oscillations abruptly set in. The oscillation period and decay time of these oscillation increase with increasing values of $\tau_{\rm f}/\tau_{\rm c}$. Asymptotically, for $t\rightarrow\infty$, $\Delta Q$ approaches the value $C \Delta V_{\rm g}$, with the capacitance $C$ given in Eq.\ (\ref{eq:C}). The approach to the asymptotic value is well described by the  terms with $n=\pm1$ in Eq.\ \eqref{eqn:opencavity3}  (red dashed lines).}
	\label{fig:chargetime}
\end{figure}

\begin{figure}[t]
	\centering
	\includegraphics[width=\linewidth]{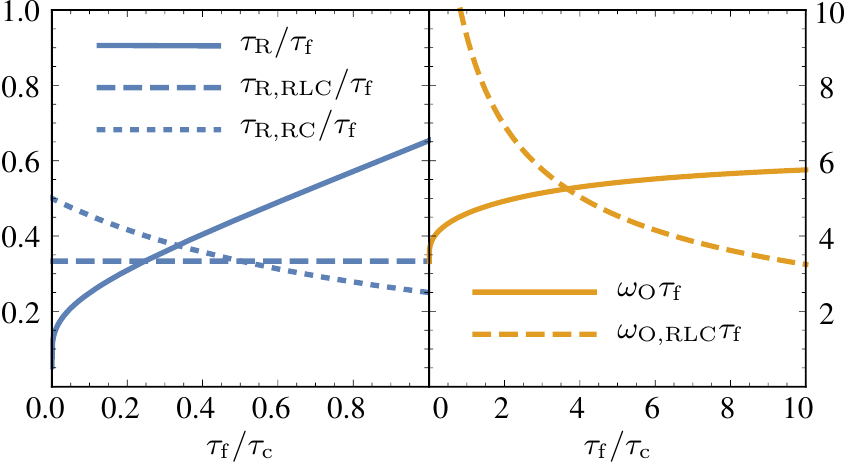}
	\caption{Relaxation time $\tau_{\rm R}$ (left panel) and oscillation frequency $\omega_{\rm O}$ (right panel), both normalized to the time of flight $\tau_{\rm f}$, for the exact solution (solid curves), for the $LRC$ circuit analogy (long dashes), and for the $RC$ circuit analogy (short dashes).}
\label{fig:times}
\end{figure}

Figure \ref{fig:chargetime} shows $\Delta Q(t)$ for $\tau_{\rm f}/\tau_{\rm c} = 0.5$, $1.5$, and $4$. (Additional curves showing $\Delta Q(t)$  in the regime of very strong/weak interaction strength, $\tau_{\rm f}/\tau_{\rm c}=20$ and $0.1$, are shown in the next Section.) The figure reveals that the system response is quite complex and that it  cannot be described by the exponential decay of a simple circuit analogy. For times $t < \tau_{\rm f}$, the behavior of the open cavity reproduces exactly the  response of an $RC$ circuit with resistance $h/e^2$ and capacitance $C_{\rm g}$. The charge displays an exponential relaxation towards the (incorrect!) asymptotic value $C_{\rm g} \Delta V_{\rm g}$ with relaxation time $\tau_{\rm c}$,
\begin{equation}
  \Delta Q(t) =  C_{\rm g} \Delta V_{\rm g} (1-e^{-t/\tau_{\rm c}}) \ \ \mbox{for $t < \tau_{\rm f}$},
  \label{eq:short}
\end{equation}
see the grey dotted line in Fig.\ \ref{fig:chargetime}. However, after time $t=\tau_{\rm f}$, $\Delta Q(t)$ changes abruptly, entering a regime of damped oscillations. These oscillations persist longer if the ratio $\tau_{\rm f}/\tau_{\rm c}$ is larger, {\it i.e.}, for increasing interaction strength. For large $t$ the oscillations are well approximated by the $n=\pm 1$ terms in the summation (\ref{eqn:opencavity3}),
\begin{equation}
  \Delta Q(t) = C \Delta V_{\rm g} - 
  \frac{2 C_{\rm g} \Delta V_{\rm g} \tau_{\rm c} e^{-t/\tau_{\rm R}}\cos(\omega_{\rm O} t + \varphi)}{|\tau_{\rm c} + \tau_{\rm f} - \tau_{\rm f} f_1|},
  \label{eq:long}
\end{equation}
with the exponential relaxation time
\begin{equation}\label{eqn:opencavity4}
  \tau_{\rm R} = \frac{\tau_{\rm c}}{\mbox{Re} f_1}
\end{equation}
and the oscillation frequency
\begin{equation}\label{eqn:opencavity5}
  \omega_{\rm O} = \frac{\mbox{Im}\, f_1}{\tau_{\rm c}}.
\end{equation}
The phase offset for the oscillations reads $\varphi = \arctan[\tau_{\rm f} \tau_{\rm c}  \tau_{\rm R} \omega_{\rm O}/(\tau_{\rm f} \tau_{\rm R} + \tau_{\rm c} \tau_{\rm R} - \tau_{\rm f} \tau_{\rm c})]$. In Fig.~\ref{fig:chargetime}, we have also included this asymptotic long-time behavior as the dashed curves \footnote{ The terms in Eq.\ (\ref{eq:at}) with $|n| > 1$, which are omitted in the approximation (\ref{eq:long}), are a factor $\sim e^{-{\rm Re}\, (f_n-f_1) t/\tau_{\rm c}}$ smaller than the $n = \pm 1$ terms kept in the approximation (\ref{eq:long}). Since $\mbox{Re}\, (f_n - f_1) \tau_{\rm R}/\tau_{\rm c} \gtrsim 0.5$ if $\tau_{\rm f}/\tau_{\rm c} \gtrsim 1$, see App.\ \ref{app:lambert}, the approximation (\ref{eq:long}) is quantitatively accurate for $t \gtrsim \tau_{\rm R}$ if $\tau_{\rm f}/\tau_{\rm c} \gtrsim 1$. For $\tau_{\rm f}/\tau_{\rm c} \sim 1$ one has $\tau_{\rm R} \sim \tau_{\rm f}$, see Fig. \ref{fig:times}, so that Eq.\ (\ref{eq:long}) is a good approximation for all $t \gtrsim \tau_{\rm f}$. For very large $\tau_{\rm f}/\tau_{\rm c}$, however, the exponential relaxation time $\tau_{\rm R} \gg \tau_{\rm f}$, and there is a large time window $\tau_{\rm f} < t \lesssim \tau_{\rm R}$ in which $\Delta Q(t)$ is neither described by Eq.\ (\ref{eq:short}) nor by Eq.\ (\ref{eq:long}). In the opposite limit of very small $\tau_{\rm f}/\tau_{\rm c}$ the terms omitted in the approximation (\ref{eq:long}) are not (yet) small if $t \sim \tau_{\rm R}$. In both cases one has to resort to the full expression (\ref{eqn:opencavity3}) to evaluate $\Delta Q(t)$.}. Asymptotically, for $t \gg \tau_{\rm R}$, $\Delta Q(t)$ relaxes to the equilibrium value $C \Delta V_{\rm g}$, the first term in Eq.~\eqref{eqn:opencavity3}.

The first kink at $t=\tau_{\rm f}$ in Fig.~\ref{fig:chargetime} and the following oscillations can be understood by inspection of the internal charge dynamics of the cavity, which will be discussed in detail in Sec.~\ref{sec:pulses}. The sharpness of the first kink derives from the sharp boundaries of the cavity, the existence of a unique time of flight $\tau_{\rm f}$ for a cavity in the quantum Hall regime, and the infinitely fast switching of the step voltage at $t=0$. In Appendix \ref{app:finiteswitching}, we illustrate how the first kink is smeared by considering a finite switching time for the step voltage, or by relaxing the assumption of sharp boundaries, a situation which could describe, for instance, non-uniform capacitive coupling to the quantum point contact as well.

In the infinite cavity limit $\tau_{\rm f}\gg\tau_{\rm c}$ damped oscillations do not occur and the mesoscopic capacitor behaves as a classical $RC$ circuit with relaxation time $\tau_{\rm R}=\tau_{\rm c}$. However, the charge evolution illustrated in Fig.~\ref{fig:chargetime} clearly shows that for finite $\tau_{\rm f}$, $\Delta Q(t)$ is not captured by the $RC$ circuit analogy, which does not allow for damped oscillations. For no time window, the relaxation time \eqref{eq:trrcshort} predicted for finite cavity sizes on the basis of the small-frequency expansion appears as a characteristic time scale of the true charge relaxation shown in Fig.\ \ref{fig:times}. 

Two recent works proposed that the charge dynamics of a mesoscopic capacitor at higher frequencies matches that of an ``$RLC$'' circuit \cite{wang2007,yin2014}. It is instructive to compare our exact solution with the predictions of a circuit of $RLC$ type. On a qualitative level there is good agreement: The step response of the $RLC$ analog displays damped oscillations towards the asymptotic value $Q(t \to \infty) = C V_{\rm g}$. Nevertheless, on a quantitative level, the relaxation time in the $RLC$ analog (see App.~\ref{app:rlc} for details)
	$\tau_{{\rm R},RLC} = \tau_{\rm f}/3$,
and the oscillation frequency
  $\omega_{{\rm O},RLC} = (1/\tau_{\rm f})\sqrt{3(1+4\,\tau_{\rm f}/\tau_{\rm c})}$
differ from the relaxation time and oscillation period obtained from our exact solution, see Eqs.\ (\ref{eqn:opencavity4}) and (\ref{eqn:opencavity5}). 

Figure \ref{fig:times} summarizes the $\tau_{\rm f}/\tau_{\rm c}$ dependence of the relaxation time $\tau_{\rm R}$ and the oscillation frequency $\omega_{\rm O}$ from the exact theory, as well as of the relaxation time and oscillation frequency from the circuit analogies. The disagreement between the exact theory and the circuit analogies is particularly apparent in the strongly interacting limit $\tau_{\rm f}\gg\tau_{\rm c}$ and confirms that low-frequency circuit analogies cannot be used to describe the non-adiabatic behavior of the interacting mesoscopic capacitor.

\begin{figure*}[t]
	\centering
	\includegraphics[width=\linewidth]{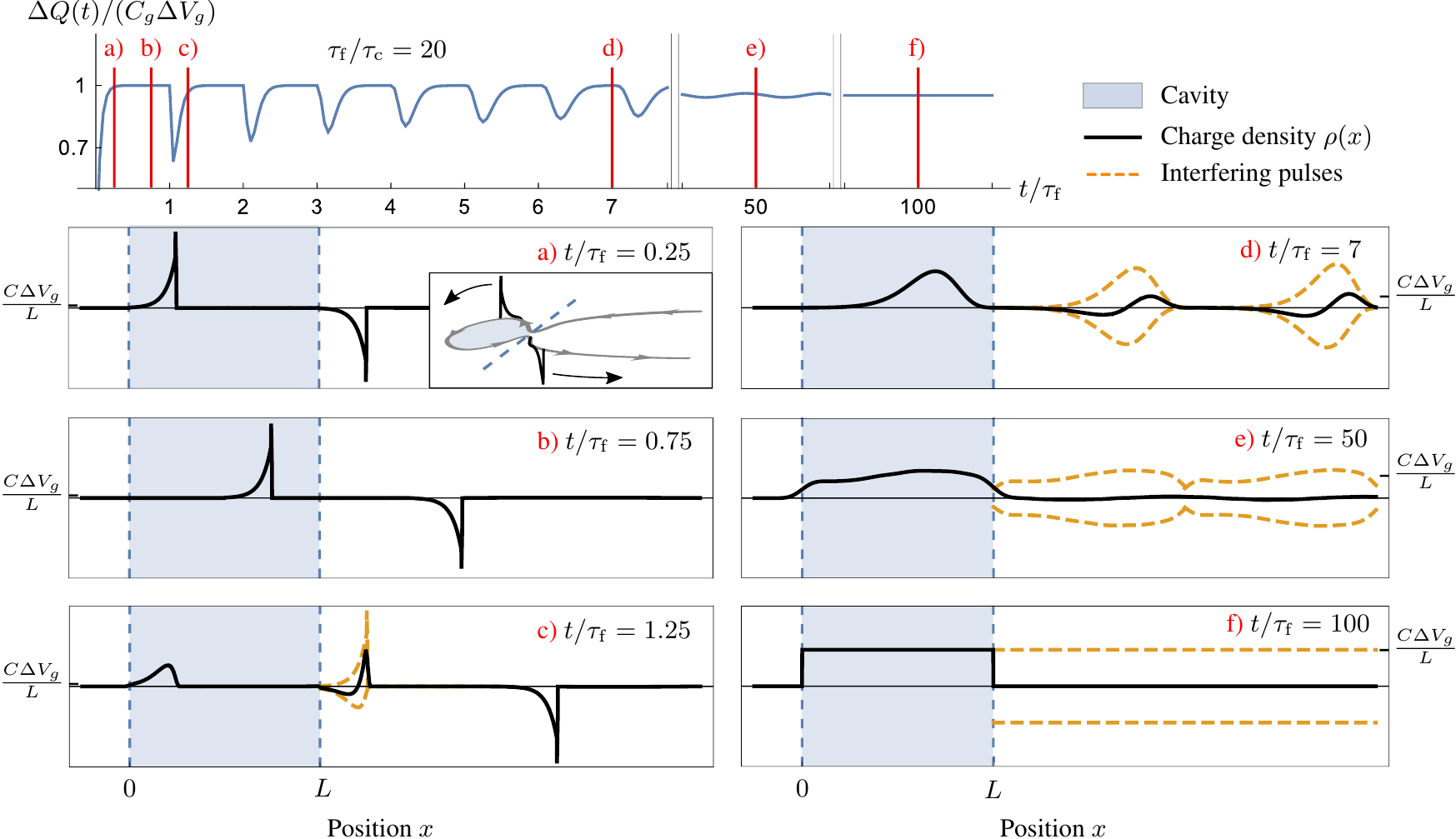}
\caption{Time-evolution of the current/charge density following a sudden gate voltage shift at time $t=0$ for large interaction strength, $\tau_{\rm f} = 20 \tau_{\rm c}$. Top: Charge response $\Delta Q(t)$ as a function of time $t$. Bottom: Series of snapshots of the current/charge density $j(x,t) = v \rho(x,t)$ at different times. In the inset of panel \textit{(a)}, the real-space representation of the mesoscopic capacitor with the profiles of the emitted charge pulses is given. The times at which the snapshots are taken are indicated by vertical dashed lines in the top panel.  Notice that the scale changes along the vertical axis in the different panels.  At time $t=0$, two charge pulses of width $\sim v \tau_{\rm c}$ and opposite sign emerge from the point contact ($a$, $b$), one pulse entering the cavity and one pulse entering the chiral edge of the bulk two-dimensional electron gas. Both pulses have a net charge approaching $C_{\rm g} \Delta V_{\rm g}$. The pulse that is emitted into the cavity returns to the point contact at time $t=\tau_{\rm f}$. As that pulse leaves the cavity, a second pulse-antipulse pair is generated ($c$), partially canceling the original charge pulse that leaves the cavity at $t = \tau_{\rm f}$. The resulting pulse exiting the cavity is the sum of the dashed profiles. The repetition of this mechanism leads to the widening and lowering of successive pulses ($d$ and $e$) (notice the change of scale between snapshots). Finally, the asymptotic configuration is attained with a charge $C \Delta V_{\rm g}$ uniformly distributed along the cavity edge ($f$). }
\label{fig:densityplots}
\end{figure*}

\section{Current dynamics}\label{sec:pulses}

To understand the origin of the kink at $t = \tau_{\rm f}$ in the time dependence of the cavity charge $\Delta Q(t)$, it is instructive to consider the current $j(x,t)$ in the chiral edge,
\begin{equation}
  j(x,t) = \frac{v e}{2 \pi} \frac{\partial \langle \phi(x,t) \rangle}{\partial x},
\end{equation}
see Eq.\ (\ref{eq:bos}). The current $j(x,t) = 0$ for $x < 0$. From the exact solution (\ref{eq:sol}), we find that
\begin{equation}
  j(x,t) = - \frac{Q(t -x/v)}{\tau_{\rm c}} +
  \frac{e^2}{h} V_{\rm g}(t-x/v)
  \label{eq:j1}
\end{equation}
for $0 < x < L$, {\it i.e.}, inside the cavity, and
\begin{align}
  j(x,t) =&\, \frac{Q(t + \tau_{\rm f}-x/v)-Q(t-x/v)}{\tau_{\rm c}} 
  \nonumber \\ &\, \mbox{} -
  \frac{e^2}{h} [V_{\rm g}(t + \tau_{\rm f}-x/v)-V_{\rm g}(t-x/v)]
  \label{eq:j2}
\end{align}
for $x > L$, {\it i.e.}, beyond the cavity. Alternatively, charge conservation gives the equivalent expression 
\begin{equation}
  j(x,t) =- \frac{\partial Q(t+\tau_{\rm f}-x/v)}{\partial t}
  \label{eq:j3}
\end{equation}
for $x>L$. [That the two expressions are equivalent follows, since equating Eqs.\ (\ref{eq:j2}) and (\ref{eq:j3}) reproduces the admittance (\ref{eqn:rccircuit1}).] For a chiral edge, the current $j(x,t)$ and the charge density $\rho(x,t)$ are proportional, $j(x,t) = v \rho(x,t)$.

The calculation of the current density profiles in response to a gate voltage step $V_{\rm g}(t) = V_{\rm g}(0) + \theta(t) \Delta V_{\rm g}$ is easily performed using the expressions for $\Delta Q(t)$ derived in the previous Section. Figure~\ref{fig:densityplots} shows the charge $\Delta Q(t)$ as well as snapshots of the current/charge density inside and outside the cavity taken at different times. We choose the ratio $\tau_{\rm f}/\tau_{\rm c} = 20$, corresponding to the limit of strong interactions. In this limit the different roles of the time scales $\tau_{\rm f}$ and $\tau_{\rm c}$ are very pronounced. The charge evolution has a ``spiked'' behavior with kinks at integer multiples of $\tau_{\rm f}$. Upon increasing time, the features decrease in amplitude and become wider, see the top panel in Fig.~\ref{fig:densityplots}. 
Similar behavior was also derived by Ngo Dinh \textit{et al.} for the time-evolution of a ``phase counting function'' describing the visibility of the interference signal in Mach-Zender interferometers with long-range Coulomb interactions  \cite{dinh12,dinh13}. This system can be described by a model which is  formally similar to the one considered here.

Snapshots $a)$ and $b)$ of the current/charge density are taken at two successive times $t<\tau_{\rm f}$ before the first kink. They show that a current pulse is emitted from the cavity within a short time $\tau_{\rm c}$ after the gate voltage quench, whereas the cavity charge $\Delta Q$ approaches the value $C_{\rm g} \Delta V_{\rm g}$ set by the new value of the gate voltage $V_{\rm g}$ on the same time scale. Importantly, the extra charge in the cavity is not localized uniformly along the edge, but constitutes a sharp charge density peak of width $\sim v \tau_{\rm c}$, traveling along the chiral edge at velocity $v$. At this stage, the time of flight $\tau_{\rm f}$ plays no role yet and the evolution of the system is fully described by the relaxation of a classical $RC$ circuit with $RC$-time $\tau_{\rm c}$. 

At time $t = \tau_{\rm f}$, the finite size of the cavity becomes apparent as the charge density pulse inside the cavity arrives at the cavity exit. The charge that accumulated inside the cavity in response to the gate voltage step now starts leaking out of the cavity, causing a kink in the charge evolution and triggering a second charge density pulse starting from the cavity entrance, see snapshot $c)$. Again, the cavity charge $\Delta Q$ approaches its asymptotic value, only slightly smaller than $C_{\rm g} \Delta V_{\rm g}$, and again the extra charge is strongly localized, although the localization profile is smoother than in panels $a)$ and $b)$. Following the structure of Eq.\ (\ref{eq:j2}) it is instructive to decompose the current pulse leaving the cavity into two pulses of opposite sign, as indicated by the dashed lines in Fig.~\ref{fig:densityplots}. 

\begin{figure}[t]
	\centering
	\includegraphics[width=\linewidth]{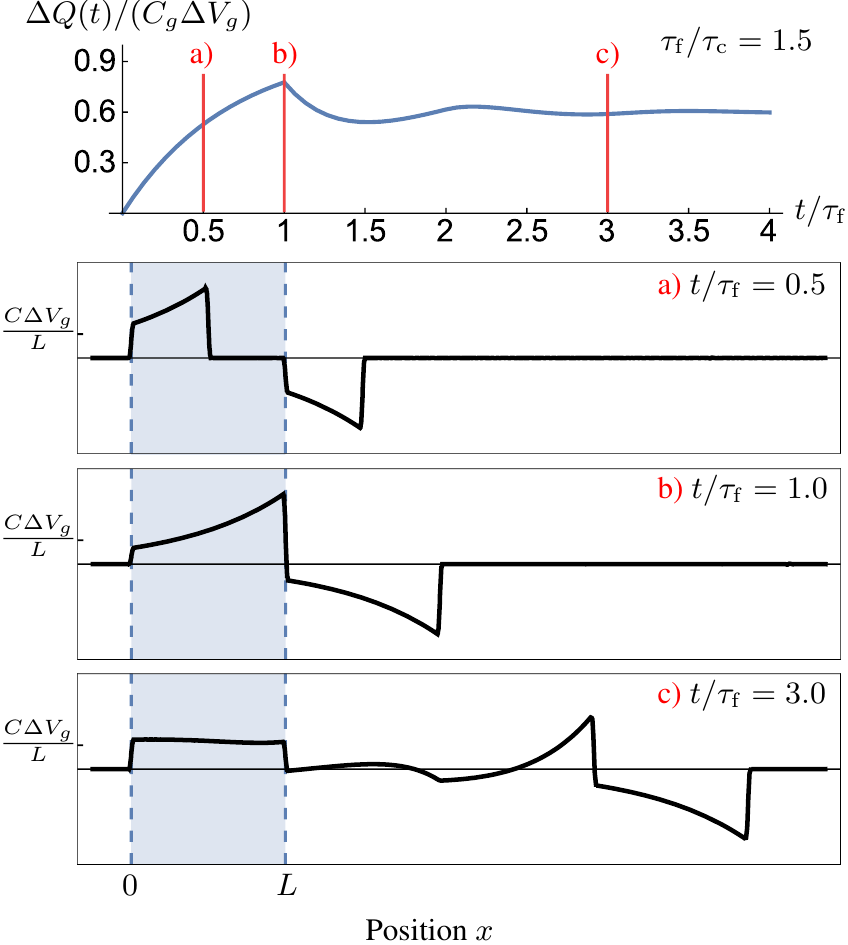}
\caption{The same as Fig.~\ref{fig:densityplots}, but for intermediate interaction strength, $\tau_{\rm f}/\tau_{\rm c}=1.5$.}\label{fig:densweak}
\end{figure}

\begin{figure}[t]
	\centering
	\includegraphics[width=\linewidth]{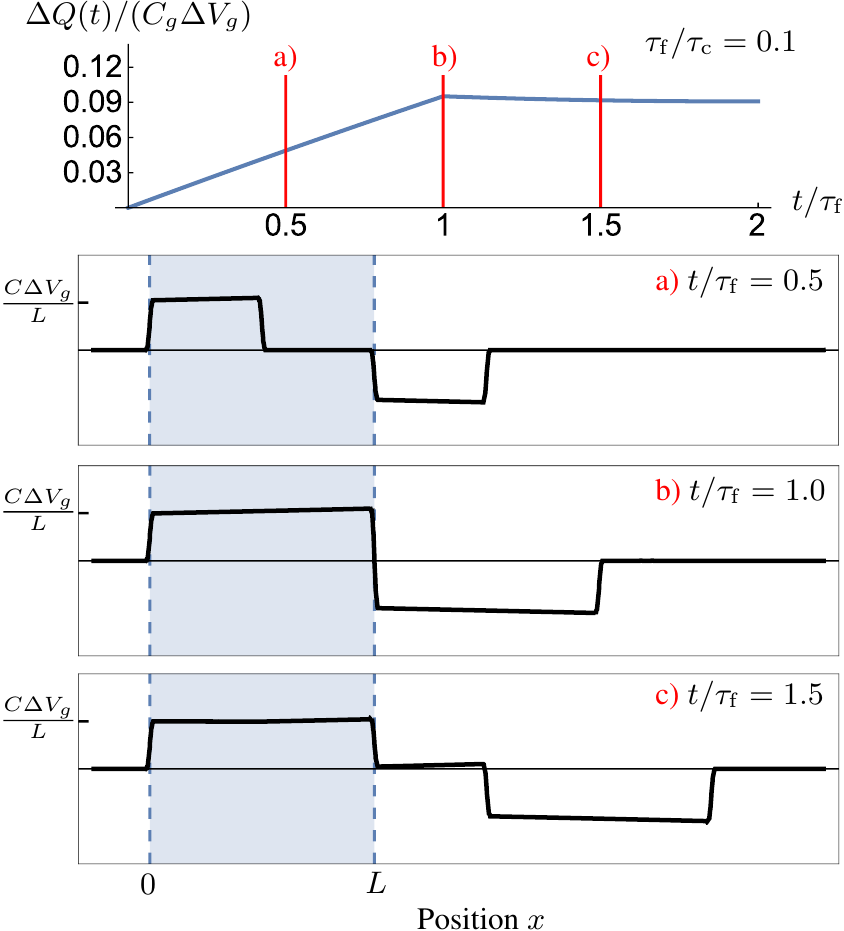}
\caption{The same as Fig.~\ref{fig:densityplots}, but for $\tau_{\rm f}/\tau_{\rm c} = 0.1$, corresponding to the weakly interacting limit. In this limit, a single flat pulse of width $v \tau_{\rm f}$ is emitted from the cavity. }\label{fig:densweak2}
\end{figure}

This procedure repeats itself, and each new density wave is wider than the previous iteration, see panels $d)$ and $e)$, because charge takes a finite and increasingly longer time to leak out of the cavity. Finally, equilibrium is attained when the complex interplay between charge leaking and filling, respectively controlled by $\tau_{\rm f}$ and $\tau_{\rm c}$, leads to a uniform configuration of the charge density within the cavity, as shown in snapshot $f)$. It is this mechanism that leads to the emergence of the quantum capacitance, such that the asymptotic value of the cavity charge is $\Delta Q(t \to \infty) = C \Delta V_{\rm g}$, where the total capacitance $C$ includes the contribution from the quantum capacitance, see Eq.\ (\ref{eq:C}).

Whereas the above discussion clarifies the separate roles of $\tau_{\rm f}$ and $\tau_{\rm c}$ in the mesoscopic capacitor in the extreme limit $\tau_{\rm f} \gg \tau_{\rm c}$, we should point out that the available experiments are in the opposite regime $\tau_{\rm f}<\tau_{\rm c}$, which is well described by self-consistent scattering theory approaches \cite{feve2007,keeling2008,moskalets2008,moskalets2013}. Figs.\ \ref{fig:densweak} and \ref{fig:densweak2} show the current response inside and outside the cavity for two more values of the ratio $\tau_{\rm f}/\tau_{\rm c}$, corresponding to interactions of intermediate strength, $\tau_{\rm f}/\tau_{\rm c} = 1.5$, and weak interactions, $\tau_{\rm f}/\tau_{\rm c} = 0.1$. In the limit of weak interactions, $\tau_{\rm f} \ll \tau_{\rm c}$, the width of the charge pulse exceeds the ``cavity size'' $L$. Instead of a sequence of charge pulses, a single almost flat pulse of width $v \tau_{\rm f}$ is emitted from the cavity. In this limit, the exponential relaxation time $\tau_{\rm R}$ becomes of the order of $\tau_{\rm f}$, in rough agreement with the scattering theory predictions and Eq. \eqref{eq:iscatt}.

We conclude this section by stressing again that the most remarkable interaction  effects are most pronounced in the $\tau_{\rm c}\ll\tau_{\rm f}$ limit, that is, for small capacitances. In this limit, the quantum point contact may contribute to the cavity capacitance as well. This additional coupling may lead to the creation of screening currents at the quantum point contact level and the emitted current, the actual measurable quantity, would have contributions which do not correspond to the inner cavity charge dynamics, as is suggested by Eq. \eqref{eq:j3}.  These  contributions may be more or less important depending on the precise design of the device, and their detailed study goes beyond the scope of this paper focusing on the role of the charging energy on the out-of-equilibrium dynamics  of the mesoscopic capacitor. Nevertheless, we show in Appendix \ref{app:finiteswitching} how our macroscopic approach can be readily extended to describe these more general situations in which the cavity does not have sharp boundaries, and how interaction screening  and capacitive  effects at the quantum point contact level may be incorporated.


\section{Backscattering Corrections}\label{sec:backscattering}

We now consider the effect of a small backscattering amplitude $r$ in the contact, described by the Hamiltonian $H_{r}$ of Eq.\ (\ref{eq:bs}). The main result of this Section is that the inclusion of  backscattering leads to a truly non-linear dependence of the charge $Q(t)$ on the gate voltage $V_{\rm g}(t)$.

\begin{figure}[t]
	\centering
	\includegraphics[width=\linewidth]{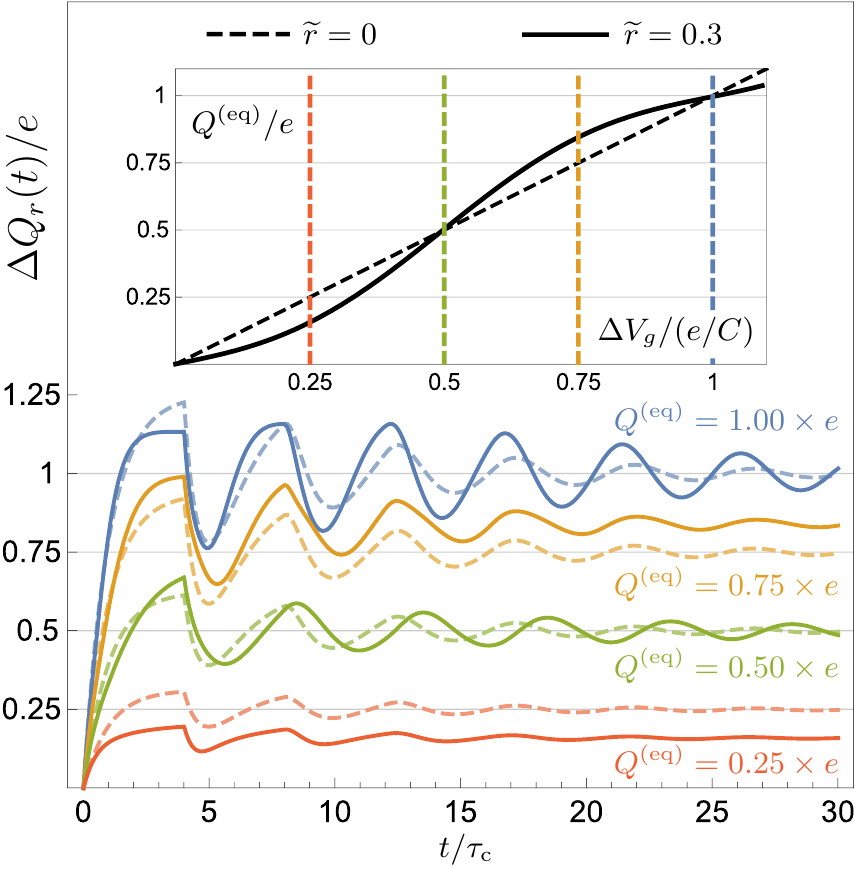}
\caption{Charge $\Delta Q_r(t)$ in response to a gate voltage step $\Delta V_{\rm g}$ for a non-ideal point contact with weak backscattering amplitude $r$. The charge $Q(0)$ at time $t=0$ is chosen to be integer-valued. Top: Asymptotic value $\Delta Q_r(t \to \infty)$ of the charge versus $\Delta V_{\rm g}$. The solid line gives the result $\Delta Q_r(t \to \infty) = C \Delta V_{\rm g}$, the dashed curve includes the first-order-in-$r$ correction to $\Delta Q_r(t \to \infty)$ in Eq. \eqref{eq:chargeeqr}. Bottom: $\Delta Q_r(t)$ versus time $t$ for $\tau_{\rm f}/\tau_{\rm c} = 4$ and $C \Delta V_{\rm g}= 0.25 e$, $0.5 e$, $0.75 e$, and $e$ --  bottom to top curves (These four values of $\Delta V_{\rm g}$ are also indicated in the top panel.).   Dashed  curves are for an ideal point contact and they are added for comparison;  solid curves include backscattering effects to first order in $r$, see Eq. \eqref{eqn:backscattering5}. All curves that include the backscattering correction are calculated for $\tilde r = 0.3$. }\label{fig:bsplot}
\end{figure}

Before discussing the non-equilibrium charge response to a time-dependent gate voltage $V_{\rm g}(t)$, we remind that already in equilibrium, weak backscattering leads to Coulomb oscillations in the equilibrium value of the charge as a function of $V_{\rm g}$ \cite{matveev1995}
\begin{equation}\label{eq:chargeeqr}
Q_{r}^{({\rm eq})} = Q^{({\rm eq})} -  \frac{e \tilde{r}}{\pi} \sin( 2\pi Q^{({\rm eq})}/e )\,,
\end{equation}
where $Q^{({\rm eq})} = C V_{\rm g}$ is the cavity charge for an ideal point contact and $\tilde{r}$ is a renormalized backscattering amplitude. The Coulomb oscillations are illustrated in the top panel of Fig. \ref{fig:bsplot}. They are precursors of the formation of charge plateaus in the limit of a cavity with tunneling point contacts. 

\begin{figure}[t]
	\centering
	\includegraphics[width=\linewidth]{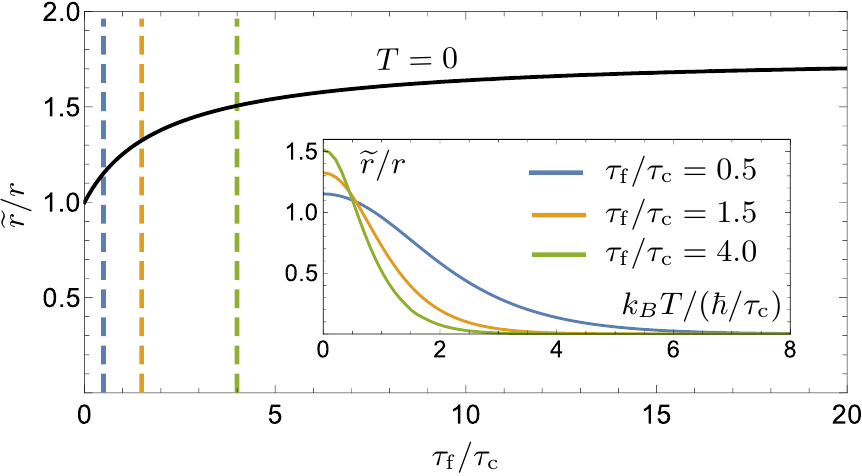}
\caption{Renormalized backscattering amplitude $\tilde r$ as a function of $\tau_{\rm f}/\tau_{\rm c}$ at zero temperature. Inset:  $\tilde r/r$ as a function of temperature for $\tau_{\rm f}/\tau_{\rm c} = 0.5$, $1.5$, and $4$, corresponding to the three values indicated in the zero-temperature plot.}\label{fig:r}
\end{figure}

We now calculate the full time-dependent charge $Q_r(t)$ in the presence of a time-dependent gate voltage $V_{\rm g}(t)$, to first order in the backscattering amplitude $r$.
To first order in the backscattering amplitude $r$, the charge $Q_r$ in the presence of backscattering can be calculated from the Kubo formula
\begin{equation}
  Q_r(t) = Q(t) + \frac{i e}{\hbar} \int^{t} dt' \langle [H_r(t'),N(t)]\rangle,
  \label{eq:Qr}
\end{equation}
where $Q(t)$ is the cavity charge in the absence of backscattering and the brackets $[\cdot,\cdot]$ denote the commutator. Since the cavity charge $N$ is linear in the bosonic fields $\phi(x,t)$, see Eq.\ (\ref{eq:N}), and the Hamiltonian $H$ in the absence of backscattering is quadratic in the bosonic fields, the average in Eq.\ (\ref{eq:Qr}) can be calculated using standard identies for operators with Gaussian fluctuations, which gives
\begin{align}
  Q_r(t) =&\, Q(t) + \frac{2 i e v r}{a} \int^{t} dt' \langle [N(t'),N(t)] \rangle \nonumber \\ &\, \mbox{} \times \sin(2 \pi \langle N(t') \rangle) e^{-2 \pi^2 (\langle N(t')^2 \rangle - \langle N(t') \rangle^2)}.
  \label{eq:Qrr}
\end{align}
Again, upon using the Kubo formula, the average of the commutator is seen to be proportional to the admittance ${\cal A}(t-t')$ in the absence of backscattering, 
\begin{equation}
  - \frac{i e^2}{\hbar} \langle [N(t'),N(t)] \rangle = {\cal A}(t-t')~~
  \mbox{if $t > t'$}.
\end{equation}
It remains to calculate the variance of the bosonic field appearing in the exponential factor. Since the non-equilibrium term proportional to $V_{\rm g}$ affects the average $\langle N(t) \rangle$ but not the fluctuations of the charge field, this factor can be obtained from the fluctuation-dissipation theorem,
\begin{align}
  \langle N(t')^2 \rangle - \langle N(t') \rangle^2 =&\,
  -\frac{i \hbar}{2\pi e^2}
  \int_{-\infty}^\infty d\omega 
  e^{-|\omega| a/v}
  \nonumber \\ &\, \mbox{} \times
  \coth(\hbar \omega/2k_{\rm B} T)
  {\cal A}(\omega),
  \label{eq:NN}
\end{align}
where the cut-off factor $e^{-|\omega| a/v}$ is compatible with the short-distance cut-off in Eq.\ (\ref{eq:bs}) (see App.\ \ref{app:backscattering} for details). The integral (\ref{eq:NN}) is logarithmically divergent for small $a$,

\begin{equation}
  \langle N(t')^2 \rangle - \langle N(t') \rangle^2 =
  - \frac{1}{2 \pi^2} \left[\ln \frac{a }{v \tau_{\rm f}} + 
  {\cal F}(T) \right],
  \label{eq:NF}
\end{equation}
where
\begin{align}
  {\cal F}(T) =\, & \int_0^\infty d\omega \, \left[\frac{1-\cos(\omega \tau_{\rm f})}{\omega}\right.
  \nonumber \\ &\left.\,-\,\frac{2\pi \hbar}{e^2}\coth\left(\frac{\hbar \omega}{2k_{\rm B} T}\right)\mbox{Im}\,{\cal A}(\omega)\right] \,
\end{align}
and the leading divergence is given by  $\int_0^\infty dy e^{-ya/v\tau_{\rm f}}\,[1-\cos(y)]/y=\frac12\ln[1+(v\tau_{\rm f}/a)^2]\rightarrow-\ln(a/v\tau_{\rm f})$. Combining Eqs.\ (\ref{eq:Qrr})--(\ref{eq:NN}) one obtains the result (\ref{eqn:backscattering5}) advertised in the introduction. In the notation of Eq.\ (\ref{eq:NF}) the expression for the renormalized backscattering amplitude $\tilde r$ takes the simple form
\begin{equation}
  \tilde r = r \frac{\tau_{\rm c}}{\tau_{\rm f}+\tau_{\rm c}}e^{{\cal F}(T)}.
  \label{eq:erre}
\end{equation}
The dependence of the renormalized backscattering amplitude at zero temperature on the ratio $\tau_{\rm f}/\tau_{\rm c}$, as well as the temperature dependence for three characteristic values of $\tau_{\rm f}/\tau_{\rm c}$ are shown in Fig.\ \ref{fig:r}.

In the presence of a finite backscattering amplitude in the quantum point contact, the response to a sudden gate voltage step becomes nonlinear in the gate voltage. The nonlinearity enters through the factor proportional to $\sin(2 \pi \langle N(t) \rangle)$ in Eq.\ (\ref{eq:Qrr}) or, equivalently, the term proportional to $\sin(2 \pi Q(t)/e)$ in Eq.\ (\ref{eqn:backscattering5}), since $Q(t)$ is proportional to $V_{\rm g}$. In Fig.\ \ref{fig:bsplot}, the effects of the non-linear corrections to the charge dynamics at finite $\tilde r$ given by Eq. \eqref{eqn:backscattering5} are illustrated and compared to the transparent limit ($\tilde r=0$). As discussed at the beginning of this Section, a finite backscattering amplitude leads to oscillations of the equilibrium charge $Q^{({\rm eq})}$ with the gate voltage $V_{\rm g}$, see Eq.\ (\ref{eq:chargeeqr}), and, hence, also to oscillations of the accumulated charge $\Delta Q_r(t \to \infty)$ with the gate voltage step $\Delta V_{\rm g}$, as shown in the top panel of Fig.\ \ref{fig:bsplot}. Concerning the approach to the asymptotic value, the evolution of the charge is not dramatically affected by the first-order backscattering correction, which essentially renormalizes relaxations times and periods of charge oscillations. However the sign of these renormalizations depends on the magnitude $\Delta V_{\rm g}$ of the voltage step. In particular, we notice that backscattering leads to longer charge relaxation times for quenches towards charge plateaus [$Q(t \to \infty)$ an integer multiple of $e$], while the period of charge oscillations increases for quenches towards charge degeneracy points [$Q(t \to \infty)-e/2$ an integer multiple of $e$].

\section{Conclusions}\label{sec:conclusion}

We studied the out-of-equilibrium behavior of the interacting mesoscopic capacitor in the large transparency limit. Our work contains the full non-equilibrium response up to first order in the reflection amplitude $r$ of the point contact connecting the cavity to the reservoir, and is tailored to the experimentally relevant case that the cavity is in the quantum Hall regime, so that electrons spend a sharply defined time $\tau_{\rm f}$ between entering and exiting the cavity. Our most important result is that for a fully transparent contact (reflection amplitude $r=0$), the charge and current response to a time-dependent change of the gate voltage $V_{\rm g}$ is strictly linear. In particular, we showed that the linear-response-in-$V_{\rm g}$ theory of Mora and Le Hur is valid irrespective of the magnitude and degree of non-adiabaticity of the gate-voltage changes $\Delta V_{\rm g}$.

The  high-frequency time-resolved response to a sudden gate-voltage step $\Delta V_{\rm g}$ allows us to clearly disentangle the two fundamental time scales in the problem. These are the ``charging time'' $\tau_{\rm c} = h C_{\rm g}/e^2$, where $C_{\rm g}$ is the geometric capacitance, and the time of flight $\tau_{\rm f}$. For times $t < \tau_{\rm f}$ the charge shows exponential relaxation with relaxation time $\tau_{\rm c}$. For $t > \tau_{\rm f}$ the charge oscillates with exponentially damped oscillations. The oscillation period and relaxation time of these oscillations can be parametrically larger than $\tau_{\rm f}$ in the limit of strong interactions. Such a scenario cannot be obtained from simple quantum circuit analogies, although the circuit analogies can capture the low-frequency dynamics correctly. Instead, the oscillations are described by the serial emission of increasingly wider charge density pulses.

To first order in the reflection amplitude $r$, we showed that the charge response becomes nonlinear in the gate voltage. To find the full non-equilibrium charge response beyond first order in $r$ and, in particular, in the weak tunneling limit remains an open problem for the case of strong Coulomb interactions and correlations.

This work paves the way towards the investigation of real-time charge emission in interacting mesoscopic devices. Our results show that a comprehensive understanding of interaction effects  unveils complex coherent dynamics triggered by interactions. Moreover, charge wave emission in the fully transparent regime  has been recently reported in Ref. \cite{freulon15} and called `edge magnetoplasmons'. Our work opens important perspectives for the experimental study of the actual pulse emitted from the cavity and how it is affected by Coulomb interactions. Moreover,  our approach can be readily extended to situations  with multiple chiral edges interacting with each other, as discussed in Refs.~\cite{bocquillon13separation,ferraro14,freulon15}, leading to qualitatively analog physics. In the case of edges in the fractional quantum Hall regime, a trivial normalization of the charge occurs in the absence of backscattering, similarly to the situation  discussed in Refs.~\cite{hamamoto2010,hamamoto2011quantum}.

We also stress that these characteristic pulses can be directly observed with available technology by ongoing experiments. Experiments in Refs.~\cite{freulon15} and~\cite{waldie15} showed the possibility to measure the shape of the emitted wave-packet by relying on Hong-Ou-Mandel experiments or by real-time modulation  of tunnel barriers respectively. 

Our results are based on an exact model in which electrons propagate in one dimension, subject to a charging interaction. Such a model is appropriate for a cavity in the quantum Hall regime, in which the electrons propagate along the chiral edge. An important feature of this model is that there is a well-defined time of flight $\tau_{\rm f}$ for electrons in the cavity. It is the sharpness of the time of flight that is responsible for the sharp features in the charge response after integer multiples of $\tau_{\rm f}$. In contrast, a cavity not in the quantum Hall regime typically has a broad distribution of dwell times, and the sharp features discussed here are not expected to appear there. 

Most experiments with mesoscopic capacitors are performed in the quantum Hall regime, because that way the emitted charge pulse follows a one-dimensional trajectory and can be more easily manipulated. Since both the time of flight $\tau_{\rm f}$ and the charging time $\tau_{\rm c}$ scale proportional with the cavity's linear size $L$, there is no a priori reason why the ratio $\tau_{\rm f}/\tau_{\rm c}$ should be large or small in such experiments. Whereas the ratio $\tau_{\rm f}/\tau_{\rm c}$ can be made small by the addition of a screening gate, we here have shown that the opposite regime $\tau_{\rm f}/\tau_{\rm c}$ of relatively strong interactions displays a very characteristic response to a gate voltage quench.

We hope that these findings will stimulate experimental efforts to also access the regime of strong interactions.


\section*{Acknowledgments}
We thank Dmitry Bagrets, Pascal Degiovanni, Gwendal F\`eve, G\'eraldine Haack, Karyn Le Hur and Christophe Mora for useful discussions. This work was supported by the DFG Priority Program 1459 (Graphene). MF was supported in part by the Swiss National Science Foundation under Division II.

\appendix

\section{Solution for the charge field $N(t)$ and its correlation function}

\label{app:backscattering}

The charge field $N(t) = [\phi(L,t)-\phi(0,t)]/2 \pi$ can be expressed in terms of the free field $\phi(0,t)$ by setting $x=L$ in Eq.\ (\ref{eq:sol}), which gives
\begin{align}
  N(t) =&\, \frac{1}{2 \pi}[\phi(0,t-\tau_{\rm f})-\phi(0,t)]
  \nonumber \\ &\, \mbox{}
  - \frac{1}{\tau_{\rm c}} \int^t dt'
  [N(t) - {\cal N}_{\rm g}(t')] s(L-v(t-t')).
\end{align}
Fourier transforming this equation to $t$ gives
\begin{equation}
  N(\omega) = {\cal A}(\omega) \left[ \frac{{\cal N}_{\rm g}(\omega)}{C_{\rm g}} - \frac{i \omega \hbar \phi(0,\omega)}{e^2}
  \right],
  \label{eq:Nomega}
\end{equation}
with ${\cal A}(\omega)$ given by Eq.\ (\ref{eqn:rccircuit1}). The inverse Fourier transform yields Eq.\ (\ref{eq:phi0eq}) of the main text, which was first derived by Mora and Le Hur for the linear response regime \cite{mora2010}.

Equation (\ref{eq:Nomega}) can be used to express the fluctuations of the charge field $N(t)$ in terms of the fluctuations of the free bosonic field $\phi(0,t)$. Using the well-known result (see, {\it e.g.}, Ref.\ \cite{vondelft1998}),
\begin{equation}
  \langle \phi(0,\omega) \phi(0,\omega') \rangle =
  \frac{2 \pi e^{-|\omega|a/v}}{\omega
  (e^{\hbar \omega/k_{\rm B} T} - 1)} \delta(\omega + \omega'),
\end{equation}
where the short-distance cut-off $a$ is the same as in Eq.\ (\ref{eq:bs}), one finds
\begin{align}
  & \langle N(\omega) N(\omega') \rangle -
  \langle N(\omega) \rangle \langle N(\omega')\rangle 
  \nonumber \\ &\ \ =
  \frac{2 \pi \omega \hbar^2{\cal A}(\omega) {\cal A}(-\omega) e^{-a|\omega|/v}}{e^2(e^{\hbar \omega/k_{\rm B} T} - 1)}
  \delta(\omega-\omega').
\end{align}
The result (\ref{eq:NN}) of the main text follows upon using the identity
\begin{equation}
  {\cal A}(\omega) {\cal A}(-\omega) 
  = \frac{e^2[A(\omega) - A(-\omega)]}{2 \pi i \hbar \omega}.
\end{equation}

\section{Lambert $W$ functions}\label{app:lambert}
The Lambert $W$ function, also called product logarithm, is the multi-valued function $z = W_n(c)$ solving the equation 
\begin{equation}
  ze^z=c.
\end{equation}
\begin{figure}
	\centering
	\includegraphics[width=\linewidth]{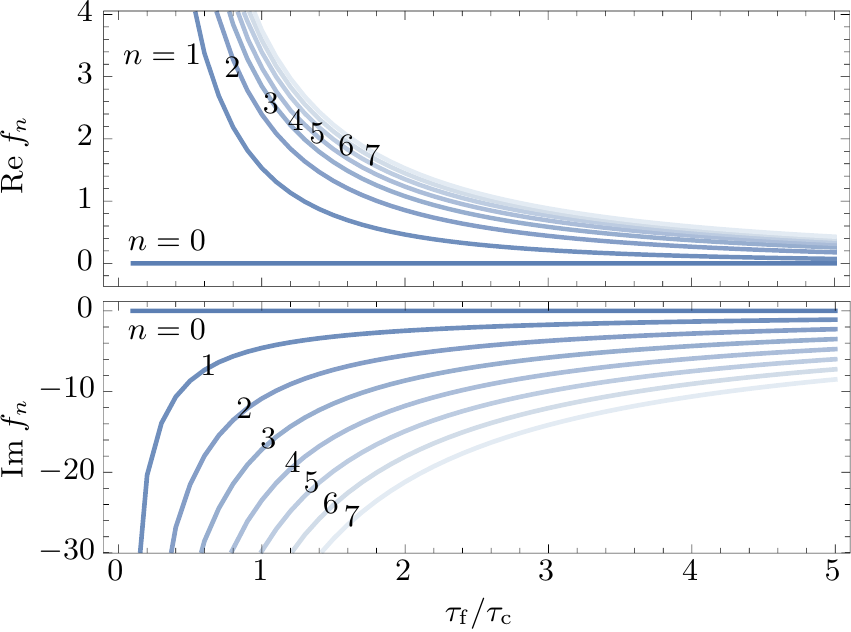}
\caption{Real and imaginary part of the function $f_n=f_{-n}^*$ defined in Eq.~\eqref{eq:fn}.  The real part is always positive and, for fixed $\tau_{\rm f}/\tau_{\rm c}$, increases with $n$. The imaginary part has opposite signs for positive and negative $n$ and also increases in absolute value with $n$.}\label{fig:lambert}
\end{figure}%
The Lambert $W$ function enters the real time representation of the admittance in Eq.~\eqref{eq:at} through the functions $f_n$ defined in Eq.~\eqref{eq:fn}. The real and imaginary parts of $f_n = f_{-n}^*$ are shown in Fig.~\ref{fig:lambert}. For fixed $\tau_{\rm f}/\tau_{\rm c}$, their absolute value increases for increasing $n$. The real part is always positive. For large values of the argument the $n$th branch $W_n$ of the Lambert $W$ function can be well approximated as
\begin{align}
  W_n(c) =&\, 2 \pi i n + \ln c 
  \nonumber \\ &\, \mbox{}
  - \ln[2 \pi i n + \ln c - \ln(2 \pi i n + \ln c)].
\end{align}

\section{Evaluation of the step response in Eq. \eqref{eqn:opencavity3}.}\label{app:numeval}

In order to obtain the long-time asymptotic value \linebreak $Q(t \to \infty) = Q^{({\rm eq})} = C \Delta V_{\rm g}$ in Eq.~\eqref{eqn:opencavity3}, we rely on the identity
\begin{equation}
	\sum\limits_{n \neq 0} \left.\frac{e^{-(z-W_n(ze^z))t/\tau_{\rm f}}}{W_n(ze^z)+1}\right|_{t \rightarrow 0^+} = \frac{z}{z+1}.
\end{equation}
The limit $t\rightarrow 0^+$ ensures the convergence of the sum and is a manifestation of the fact that the time $t$ in Eq.\ \eqref{eq:at} is always positive.\\[0.5em]

\begin{figure}[t]
	\centering
	\includegraphics[width=\linewidth]{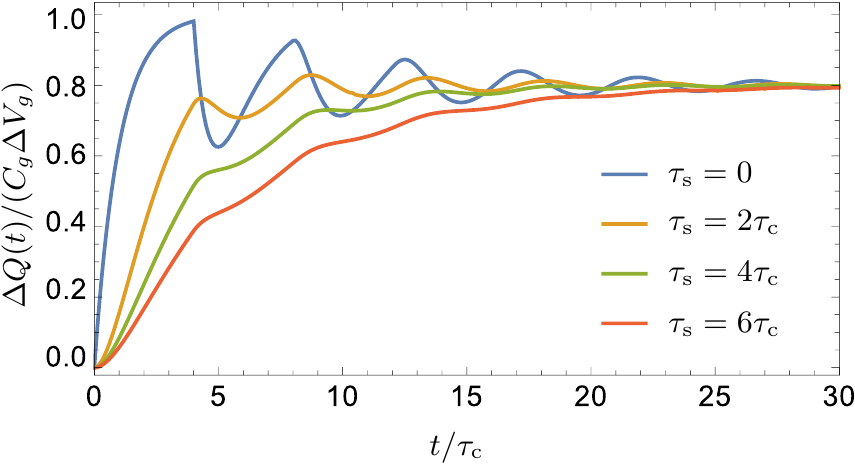}
	\caption{Charge response of an open cavity with $\tau_{\rm f} = 4\tau_{\rm c}$ to step voltages with different switching times $\tau_{\rm s}$.}
	\label{fig:finiteswitching}
\end{figure}

\section{Step response with finite switching time and extension to non-uniform capacitive coupling}
\label{app:finiteswitching}

In Fig.~\ref{fig:chargetime}, we show the response of the open cavity to a step voltage $V_{\rm g}(t) = V_{\rm g} \theta(t)$. Since this voltage is discontinuous at $t=0$, the step response has kinks at $t=0$ and $t=\tau_{\rm f}$, where the kink at $\tau_{\rm f}$ is due to the first discontinuous charge density pulse leaving the cavity (see Fig.~\ref{fig:densityplots}).

In order to demonstrate that these kinks vanish as we introduce a finite switching time $\tau_s$, we show the step response for a voltage
\begin{equation}
  V_{\rm g}(t) = V_{\rm g}(0) = \Delta V_{\rm g} \theta(t) (1-e^{-t/\tau_{\rm s}}).
\end{equation}
The results in Fig.~\ref{fig:finiteswitching} show that the kinks indeed vanish if the switching time is larger than the charge relaxation time. The oscillations in the charge response persist as long as $\tau_{\rm s} \lesssim \tau_{\rm f}$.

Similar conclusions for the charge emitted from the mesoscopic capacitor can be made if we relax the assumption of a cavity with sharply defined boundaries, in which, for instance, also the quantum point contact may lead to capacitive coupling to both the cavity and the outer edge. In this case, screening currents may be generated at the quantum point contact level, which also lead to the kink smearing, but also to corrections to Eqs. \eqref{eq:j2} and \eqref{eq:j3}. The most general capacitive coupling that also includes interactions between electrons inside and outside of the cavity, is obtained by replacing Eq.~\eqref{eq:charging} with
\begin{equation}\label{eq:intgen}
H_{\rm int}=\int dx dy U(x,y) \rho(x)\rho(y) - \int dx \gamma(x,t) \rho(x)\,,
\end{equation}
in which $\rho(x)=:\psi^\dagger(x)\psi(x):$ is the electron density in the chiral edge, $U(x,y)$ encodes electron-electron interactions and $\gamma(x,t)$ is the electrostatic time-dependent potential which is applied on the edge state by metallic gate contacts. The equation of motion \eqref{eq:eqm} readily generalizes to  
\begin{equation}\label{eq:eqmgen}
  \frac{\partial \phi(x,t)}{\partial t} =-v_F \frac{\partial \phi(x,t)}{\partial x} -\int dy\frac{U(x,y)}{\pi\hbar} \frac{\partial \phi(y,t)}{\partial y}+\frac{\gamma(x,t)}\hbar\,,
\end{equation}
which has a solution analog to Eq. \eqref{eq:sol}
\begin{align}\label{eq:solgen}
 \phi(x,t) =&\, \phi(0,t-x/v) +\int ^t dt'\left[\frac{\gamma(x-v(t-t'),t)}\hbar\right.\nonumber 
  \\ &\quad-\left.\int dy \frac{U(x-v(t-t'),y)}{\pi\hbar}\frac{\partial \phi(y,t')}{\partial y}\right]\,.
\end{align}
The modifications to the results presented in the main text, can be appreciated by considering 
\begin{align}
U(x,y)&= s(x) s(y)\frac{e^2}{2C_{\rm g}}+f(x,y)\,, \\
 \gamma(x,t)&=s(x)eV_{\rm g }(t)+g(x,t)\,.
\end{align}
The function $s(x)$ is given in Eq.~\eqref{eq:s} and Eq.~\eqref{eq:charging} is recovered by setting both $f(x,y)$ and $g(x,t)$ to zero in Eq.~\eqref{eq:intgen}. The function $f(x,y)$ can at the same time describe screening effects, leading to sound velocity renormalization if $f(x,y)$ is short ranged, and capacitive coupling through the quantum point contact if  $f(x,y)$ has a spatial long-range support outside the spatial window $0<x,y<L$. The function $g(x,t)$ being non zero outside this window describes electric potential variations outside of the cavity, which may equally occur. The presence of these terms leads corrections to Eqs. \eqref{eq:j2} and \eqref{eq:j3} of the form $j'(x,t)=j(x,t)+\delta j(x,t)$ in which 
\begin{align} \label{eq:japp}
  \delta j(x,t) =&\frac{ev}{2\pi}\frac\partial{\partial x}\int ^t dt'\left[\frac{g(x-v(t-t'),t)}\hbar\right. 
  \\ &\quad-\left.\int dy \frac{f(x-v(t-t'),y)}{\pi\hbar}\frac{\partial \phi(y,t')}{\partial y}\right]\,. \nonumber
\end{align}
This contribution describes screening currents generated out of the cavity, either by electric potential variations out of the cavity (first term) or by capacitive coupling between electrons inside and outside of  the cavity through the quantum point contact (second term).  Beyond
 a smearing of the current signal and the kinks in the charge dynamics of the cavity~--~analogous to that occurring for the charge response with finite switching times given in Fig.~\ref{fig:chargetime}~--~both these currents  lead in general to corrections of Eq. \eqref{eq:j2} and \eqref{eq:j3}, meaning that the measured current may not  fully correspond  to the internal charge dynamic of the cavity. The relevance of this signal poisoning depends on the precise design of the device and it can be readily described with the current approach.


\section{Comparison with $RLC$ circuit.}\label{app:rlc}

The expansion of the admittance Eq. \eqref{eqn:rccircuit1} to second order in frequency matches the one of a classical $RLC$ circuit  
\begin{align}\label{eq:rlc}
  \frac{\Delta Q(\omega)}{\Delta V_{\rm g}(\omega)} &= \frac{C}{1-i\omega R_qC - LC \omega^2} \\
 &= C+R_qC^2 i\omega+\big(C^2  L-C^3R_q^2\big)\omega^2 + {\cal O}(\omega^3), \nonumber
\end{align}
with $C$ given by Eq.\ (\ref{eq:C}), $R_{\rm q} = h/2e^2$, and \cite{wang2007,yin2014}
\begin{equation}\label{eq:crl}
  L = CR_q^2 \frac{\tau_{\rm f}+\tau_{\rm c}}{3 \tau_{\rm c}}\,.
\end{equation}
Relying on the first line of Eq.\ \eqref{eq:rlc}, before low-frequency expansion, the charge evolution after a quench of $V_{\rm g}$ leads  to damped oscillations. The relaxation time $\tau_{{\rm R},RLC}$ and oscillation frequency $\omega_{{\rm O},RLC}$ are readily extracted from the poles of Eq. \eqref{eq:rlc}, which leads to the values mentioned in Sec.\ \ref{sec:step} of the main text.


\bibliographystyle{apsrev4-1}
\bibliography{biblio}

\end{document}